\documentclass[10pt,letterpaper]{article}
\usepackage[top=0.85in,left=2.75in,footskip=0.75in]{geometry}
\usepackage{comment}

\usepackage{algpseudocode}
\usepackage{algorithm}
\usepackage[inline]{enumitem}
\usepackage{longtable}

\usepackage{amsmath,amssymb}

\usepackage{changepage}

\usepackage[utf8x]{inputenc}

\usepackage{textcomp,marvosym}

\usepackage{cite}

\usepackage{nameref,hyperref}

\usepackage[right]{lineno}

\usepackage{microtype}
\DisableLigatures[f]{encoding = *, family = * }

\usepackage[table]{xcolor}

\usepackage{array}

\usepackage{multirow}
\usepackage{booktabs}
\usepackage{makecell}

\newcolumntype{+}{!{\vrule width 2pt}}

\newlength\savedwidth

\newcommand\thickhline{\noalign{\global\savedwidth\arrayrulewidth\global\arrayrulewidth 2pt}%
\hline
\noalign{\global\arrayrulewidth\savedwidth}}

\raggedright
\usepackage[aboveskip=1pt,labelfont=bf,labelsep=period,justification=raggedright,singlelinecheck=off]{caption}

\makeatletter
\renewcommand{\@biblabel}[1]{\quad#1.}
\makeatother

\usepackage{lastpage,fancyhdr,graphicx}
\usepackage{epstopdf}
\fancyheadoffset[L]{2.25in}
\fancyfootoffset[L]{2.25in}
\begin{document}
\vspace*{0.2in}

\begin{flushleft}
{\Large
\textbf\newline{Criminal Networks Analysis in Missing Data scenarios through Graph Distances} 
}
\newline
\\
Annamaria Ficara\textsuperscript{1,3\Yinyang},
Lucia Cavallaro\textsuperscript{2\Yinyang*},
Francesco Curreri\textsuperscript{1,3\Yinyang},
Giacomo Fiumara\textsuperscript{3\ddag},
Pasquale De Meo\textsuperscript{4\ddag},
Ovidiu Bagdasar\textsuperscript{2\ddag},
Wei Song\textsuperscript{5\ddag},
Antonio Liotta\textsuperscript{6\ddag}
\\
\bigskip
\textbf{1} DMI Department, University of Palermo, Palermo, Italy
\\
\textbf{2} School of Computing and Engineering, University of Derby, UK
\\
\textbf{3} MIFT Department, University of Messina, Messina, Italy
\\
\textbf{4} DICAM Department, University of Messina, Messina, Italy
\\
\textbf{5} College of Information Technology, Shanghai Ocean University, Shanghai, China 
\\
\textbf{6} Faculty of Computer Science, University of Bozen-Bolzano, Bozen-Bolzano, Italy
\\
\bigskip

%
%
\Yinyang These authors contributed equally to this work.

\ddag These authors also contributed equally to this work.



\textpilcrow Membership list can be found in the Acknowledgments section.

* l.cavallaro@derby.ac.uk
\end{flushleft}
\section*{Abstract}
Data collected in criminal investigations may suffer from:
\begin{enumerate*}[label=(\roman*)]
\item incompleteness, due to the covert nature of criminal organisations;
\item incorrectness, caused by either unintentional data collection errors and intentional deception by criminals;
\item inconsistency, when the same information is collected into law enforcement databases multiple times, or in different formats.
\end{enumerate*}
In this paper we analyse nine real criminal networks of different nature (i.e., Mafia networks, criminal street gangs and terrorist organizations) in order to quantify the impact of incomplete data and to determine which network type is most affected by it. The networks are firstly pruned following two specific methods: \begin{enumerate*}[label=(\roman*)]
\item random edges removal, simulating the scenario in which the Law Enforcement Agencies (LEAs) fail to intercept some calls, or to spot sporadic meetings among suspects; 
\item nodes removal,
that catches the hypothesis in which some suspects cannot be intercepted or investigated. 
\end{enumerate*} 
Finally we compute spectral (i.e., Adjacency, Laplacian and Normalised Laplacian Spectral Distances) and matrix (i.e., Root Euclidean Distance) distances between the complete and pruned networks, which we compare using statistical analysis. 
Our investigation identified two main features: first, the overall understanding of the criminal networks remains high even with incomplete data on criminal interactions (i.e., 10\% removed edges); second, removing even a small fraction of suspects not investigated (i.e., 2\% removed nodes) may lead to significant misinterpretation of the overall network.


\section*{Introduction}

Criminal organizations are groups operating outside the boundaries of the law, which make illegal profit from providing illicit goods and services in public demand and whose achievements come at the cost of other people, groups or societies~\cite{Finckenauer2005}. Organised crime is referred to different terms including \textit{gangs}~\cite{thrasher2013gang}, \textit{crews}~\cite{adler1993wheeling}, \textit{firms}~\cite{reuter1983disorganized}, \textit{syndacates}~\cite{reuter1983disorganized}, or \textit{Mafia}~\cite{Morselli2008}. In particular, Gambetta \cite{gambetta1996sicilian} defines Mafia as a ``territorially based criminal organization that attempts to govern territories and markets'' and he identifies the one located in Sicily as the \textit{original Mafia}.

Whatever term is used to call the organised crime, this involves relational traits. For this reason, scholars and practitioners are increasingly adopting a Network Science Analysis (SNA) perspective to explore criminal phenomena~\cite{Campana2016}.
SNA algorithms can produce relevant measurements and parameters describing the role and importance of individuals within criminal organizations, and SNA has been used to identify leaders within a criminal organization~\cite{Johnsen2018IdentifyingCI}
and to construct crime prevention systems~\cite{Calderoni2020}.

Over the last decades, SNA has been employed increasingly by Law Enforcement Agencies (LEAs). This increasing interest from law enforcement is due to SNA's ability to identify mechanisms that are not easily discovered at first glance~\cite{morselliglance}. 

SNA relies on real datasets used as sources which allow to build networks that are then examined~\cite{Duijn2014, Rostami2015, Robinson2018, Villani2019, Ficara2020, Calderoni2020, Cavallaro2020, Cavallaro2021IoT}. 
However, the collection of complete network data describing the structure and activities of a criminal organization is difficult to obtain.

In a criminal investigation, the individuals subjected to LEAs enquiries may attempt to shield sensible information. Investigators then have to rely on alternative methods and exercise special investigative powers allowing them to gather evidence covertly from sources including phone taps, surveillance, archives, informants, interrogations to witnesses and suspects, infiltration in criminal groups.
Despite significant advantages, such sources may also have a number of drawbacks.

Also, while some individuals providing information during investigations are reliable, others might attempt to deceive the investigations with the aim to protect themselves, their associates, or to achieve a specific goal. For instance, if actors are aware of being phone-tapped, they are more likely to avoid to discuss of self-incriminating evidence.
While the transcripts of discussions between unsuspecting actors can be considered more reliable, the information collected from taps must still be verified against other official records related to the case. This is required since conversations among criminals often involve lies or codes concealing the true nature of the message~\cite{campanavarese2013}. 
Moreover, if police misses surveillance targets, central actors may not appear with their actual role in the data, simply because their phones end up not being tapped~\cite{Morselli2008}.

While the police seeks to validate the content of phone-taps, the offenders may also check themselves whether the information received from the police during conversations is accurate. Longer investigations and surveillance tend to eventually expose subtle lies. On the other side, datasets may change with time, due to the variable status of suspects, or to new information being collected. The problem of actors lying is extended to data collected through questionnaires or interviews as well. Information collected from interrogations may not be reliable, with the risk of interviewees downplaying or amplifying their real role, or simply not being representative of the broader group. 

Police decisions may even impact the design of an investigation. LEAs normally start with some suspected individuals, and then expand their reach by adding further actors. Not all the individuals linked to the central actors are automatically added, as the investigation of all active criminal groups is not possibile due to limited resources. Prosecution services must prioritise the groups on which evidence gathering is easier. Hence, groups operating under the police radar may be absent from the data collected, and this may generate heavily distorted inferences about the network structure~\cite{snaresearch}.

Incompleteness and incorrectness of criminal network data is then inevitable. This is due to investigators dealing with data of different quality and because in SNA there is currently no standard method to account for such degrees of reliability.

LEAs often have to process lots of data, most of which is of little value. When large volumes of raw data are collected from multiple sources, the risk of inconsistency is also higher.
The identification of relevant and important information from datasets where this is mixed with irrelevant or unreliable information, is referred to as the {\em signal and noise} problem.
Analytical techniques used in intelligence should then be able to cope with large datasets and to effectively distinguish the signal from noise.

In summary, the data collected in criminal investigations regularly suffers from:
\begin{itemize}
\item {\em Incompleteness}, caused by the covert nature of criminal networks;
\item {\em Incorrectness}, caused by either unintentional errors in data collection or intentional deception by criminals;
\item {\em Inconsistency}, when records of the same actors are collected into LEA databases multiple times and not necessarily in a consistent way. This way, the same actor may show up in a network as different individuals.
\end{itemize}

Criminal networks are very dynamic, as they constantly change over time. New data or even different data collection methods are necessary to cover longer time spans~\cite{SPARROW1991251}.
Another problem specific to SNA used for criminal networks lies in data processing. 
Often, actors are represented by nodes, and their associations or interactions by links. However, there is no SNA standard methodology for transforming the raw data and the process depends on the subjective judgement of the analyst. This may have to decide whom to include or exclude from the network, when boundaries are ambiguous~\cite{SPARROW1991251}. Also, data conversion is often a labor-intensive and time-consuming process.

An interesting application of SNA is to compare networks by finding and quantifying similarities and differences \cite{Squartini2015, Peixoto2018, Newman2018}. Network comparison requires measures for the distance between graphs, which can be done using multiple metrics. That is a non-trivial task which involves a set of features that are often sensitive to the specific application domain. A few literature reviews on the most common graph comparison metrics are available~\cite{Soundarajan2014, Emmert2016, Donnat2018, Tantardini2019}. In~\cite{Cavallaro2021}, such distance measures were exploited to quantify how much artificial, but also realistic models can represent real criminal networks. 

In this work, we adopt a SNA approach to assess the impact of incomplete data in a criminal network. Our aim is to quantify how much information on the criminal network is required, so that the accuracy of investigations is not affected. Specifically, we analyse nine real criminal networks of different nature, which are the result of different investigative operations over Mafia networks, criminal street gangs and terrorist organizations. 
To quantify the impact of incomplete data and to determine which network suffers mostly from it, we adopt the following strategies:
\begin{enumerate}
    \item We pruned input networks by means of two specific methods, namely:
    {\em random edges removal} and {\em random nodes removal}, which reflect the most common scenarios of missing data arising in real investigations. 
    \item We calculated the distance between the original (defined as complete as a reference) network and its pruned version.
\end{enumerate}

\section*{Materials and methods}
This section presents basic graph theory definitions and the distance metrics used for comparing two graphs. We also describe the datasets used in our experimental analysis, as well as the protocol followed to run our analysis.

\subsection*{Background}
\subsubsection*{Graph properties}

A {\em network} (or {\em graph}) $G = \langle N, E\rangle$ consists of two finite sets $N$ and $E$~\cite{barabasi2016network}. The set $N=\{1,\dots,n\}$ contains the {\em nodes} (or vertices, actors), and $n$ is the \textit{size} of the network, while the set $E \subseteq N \times N$ contains the {\em edges} (or links, ties) between the nodes.

A network is called {\em undirected} if all its edges are bidirectional. If the edges are defined by ordered pairs of nodes, then the network is called {\em directed}. 
If an edge $(i, j)$ with $i,j\in N$ is {\em weighted}, then a positive numerical weight $w_{ij}$ is associated; the {\em unweighted} edges have their weight set to the default value $w_{ij}=1$.

Given an undirected network $G$, two nodes $i,j \in N$ are {\em connected} if there is a {\em path} from $i$ to $j$: here a path $p$ is defined as a sequence of nodes $i_0, i_1, \ldots, i_k$ such that each pair of consecutive nodes is connected through an edge. The number of edges in a path $p$ starting at node $i$ and ending at node $j$ is called {\em path length}. While there may be several paths from the node $i$ to the node $j$, we are usually interested in the {\em shortest paths} (i.e., those with the least number of edges), whose length defines the {\em distance} $d_{ij}$ between $i$ and $j$. Of course, in undirected networks we have $d_{ij} = d_{ji}$.

A graph $G$ is called {\em connected} if every pair of nodes in $G$ is connected, and {\em disconnected} otherwise. If a network is disconnected, it fragments into a collection of connected subnetworks, each of them called {\em components}.

Based on the number of edges $m$, a graph is called {\em dense} if $m$ is of the same order of magnitude as $n^2$, or {\em sparse} if $m$ is of the same order of magnitude as $n$. The {\em density} $\delta$ of an undirected graph is defined as
\begin{eqnarray}
\delta=\frac{2 \lvert E \rvert}{\lvert V \rvert (\lvert V \rvert-1)} = \frac{2m}{n(n-1)},
\end{eqnarray}
that is the total number of edges over the maximum possible number of edges.

The degree $k_i$ of the node $i$ represents the number of adjacent edges, while the degree distribution $p_k$ provides the probability that a randomly selected node in the graph has degree $k$. Given a graph of $n$ nodes, $p_k$ is the normalised histogram given by
\begin{eqnarray}
p_k=\frac{n_k}{n},
\end{eqnarray}
where $n_k$ is the number nodes of degree $k$.

The degree $k_i$ allows to compute the {\em clustering coefficient} $C_i$ of a node $i$~\cite{Ficara2021IoT}, which captures the degree to which the neighbors of the node $i$ link to each other, given by
\begin{eqnarray}
C_i=\frac{2L_i}{k_i(k_i-1)},
\end{eqnarray}
where $L_i$ represents the number of links between the $K_i$ neighbors of node $i$. The average of $C_i$ over all nodes defined the average clustering coefficient $\langle C_i \rangle$,
measuring the probability that two neighbors of a randomly selected node link to each other.

Given a pair of graphs, say $G_1$ and $G_2$, we are often interested in defining a measure of similarity (or, equivalently, distance) between them. In what follows we review some methods one can use to compute the distance of two graphs. 

\subsubsection*{Spectral distances}
Spectral distances allow to measure the structural similarity between two graphs starting from their spectra. The spectrum of a graph is widely used to characterise its properties and to extract information from its structure.

The most common matrix representations of a graph are the adjacency matrix $A$, the Laplacian matrix $L$ and the normalised Laplacian $\mathcal{L}$.

Given a graph $G$ with $n$ nodes, its adjacency matrix $A$ is an $n \times n$ square matrix denoted by $A =(a_{ij})$, with $1\leq i,j\leq n$, where $a_{ij} = 1$ if there exists an edge joining nodes $i$ and $j$, and $a_{ij} = 0$ otherwise.

For undirected graphs the adjacency matrix is symmetric, i.e., $a_{ij}$=$a_{ji}$.

The degree matrix $D$ is a diagonal matrix where $D_{ii} = k_i$ and $D_{ij} = 0$ for $i\neq j$.
\begin{eqnarray}
{D_{ij}} =
\begin{cases}
k_i & \text{if } i=j \\
0             & \text{otherwise}
\end{cases}
\end{eqnarray}
The adjacency matrix and the degree matrix are used to compute the combinatorial Laplacian matrix $L$, which is an $n \times n$ symmetric matrix defined as
\begin{eqnarray}
\label{eq:lmatrix}
L = D - A.
\end{eqnarray}
The diagonal elements $L_{ii}$ of $L$ are then equal to the degree $k_i$ of the node $i$, while off-diagonal elements $L_{ij}$ are $-1$ if the node $i$ is adjacent to $j$ and 0 otherwise.
A normalised version of the Laplacian matrix, denoted as $\mathcal{L}$, is defined as 
\begin{eqnarray}
\label{eq:nlaplacian}
\mathcal{L} = D^{-\frac{1}{2}} LD^{-\frac{1}{2}},
\end{eqnarray}
where the diagonal matrix $D^{-\frac{1}{2}}$ is given by
\begin{eqnarray}
\label{eq:dmatrix}
{D^{-\frac{1}{2}}_{i,i}} =
\begin{cases}
\frac{1}{\sqrt{k_i}} & \text{if } k_i \neq 0\\
0       & \text{otherwise.}
\end{cases}
\end{eqnarray}

The spectrum of a graph consists of the set of the sorted eigenvalues of one of its representation matrices. The sequence of eigenvalues may be ascending or descending depending on the chosen matrix. The spectra derived from each representation matrix may reveal different properties of the graph. The largest eigenvalue absolute value in a graph is called the \textit{spectral radius} of the graph.
If $\lambda^A_k$ is the $k^{th}$ eigenvalue of the adjacency matrix $A$, then the spectrum is given by the descending sequence 
\begin{eqnarray}
\label{eq:spectrum1}
\lambda^A_1 \geq \lambda^A_2 \geq \dots \geq \lambda^A_n.
\end{eqnarray}
If $\lambda^L_k$ is the $k^{th}$ eigenvalue of the Laplacian matrix $L$, such eigenvalues are considered in ascending order so that
\begin{eqnarray}
\label{eq:spectrum2}
0=\lambda^L_1 \leq \lambda^L_2 \leq \dots \leq \lambda^L_n.
\end{eqnarray}
The second smallest eigenvalue of the Laplacian matrix of a graph is called its \textit{algebraic connectivity}.
Similarly, if we denote the $k^{th}$ eigenvalue of the normalised Laplacian matrix $\mathcal{L}$ as $\lambda^{\mathcal{L}}_k$, then its spectrum is given by
\begin{eqnarray}
\label{eq:spectrum3}
0=\lambda^{\mathcal{L}}_1 \leq \lambda^{\mathcal{L}}_2 \leq \dots \leq \lambda^{\mathcal{L}}_n.
\end{eqnarray}

The {\em spectral distance} between two graphs is the euclidean
distance between their spectra~\cite{Wilson2008}.
Given two graphs $G$ and $G^{\prime}$ of size $n$, with their spectra respectively given by the set of eigenvalues $\lambda_i$ and $\lambda_{i'}$, their spectral distance, according to the chosen representation matrix, is computed as follows by the formula
\begin{eqnarray}
\label{eq:spectdistance}
{d(G,G')} = \sqrt{\sum_{i=0}^n (\lambda_i - \lambda_{i'})^2}.
\end{eqnarray}

Based on the chosen representation matrix and consequently its spectrum, the most common spectral distances are the adjacency spectral distance $d_A$, the Laplacian spectral distance $d_L$ and the normalised Laplacian spectral distance $d_{\mathcal{L}}$.

If the two spectra are of different sizes, the smaller graph is brought to the same cardinality of the other by adding zero values to its spectrum. In such case, only the first $k \ll n$ eigenvalues are compared.
Given the definitions of spectra of the different matrices, the adjacency spectral distance $d_A$ compares the
largest $k$ eigenvalues, while $d_{L}$ and $d_{\mathcal{L}}$ compare the smallest $k$ eigenvalues. This determines the scale at which the graphs are studied, since comparing the higher eigenvalues allows to focus more on global features, while the other two allow to focus more on local features.

\subsubsection*{Matrix distances}
Another class of distances between graphs is the matrix distance~\cite{Wills2020}. A matrix of pairwise distances $d_{ij}$ between nodes on the single graph is constructed for each as
\begin{eqnarray}
\label{eq:distmat}
M_{ij}=d_{ij}.
\end{eqnarray}

While most common distance $d$ is the shortest path distance, other measures can also be used, such as the effective graph resistance or variations on random-walk distances.
Such matrices provide a signature of the graph characteristics and carry important structural information. Matrices $M$ are then compared using some norm or distance.

Given two graphs $G$ and $G^{\prime}$, with $M$ and $M^{\prime}$ being their respective matrices of pairwise distances, the matrix distance between the $G$ and $G^{\prime}$ is introduced as:
\begin{eqnarray}
\label{eq:genmatdist}
d(G,G^{\prime})= \|M-M^{\prime}\|,
\end{eqnarray}
where $\|.\|$ is a norm to be chosen. If the matrix used is the adjacency matrix $A$, the resulting distance is called \textit{edit distance}.

The similarity measure used in this work is called \textsc{DeltaCon}~\cite{koutra2013}. It is based on the root euclidean distance $\mathrm{d_{rootED}}$, also called \textit{Matsusita difference}, between matrices $S$ created from the fast belief propagation method of measuring node affinities.

The \textsc{DeltaCon} similarity is defined as
\begin{eqnarray}
\label{eq:deltaconsim}
sim_{DC}(G,G^{\prime})=
\frac{1}{1+\mathrm{d_{rootED}}(G,G^{\prime}),}
\end{eqnarray}
where $\mathrm{d_{rootED}}(G,G^{\prime})$ is defined as
\begin{eqnarray}
\label{eq:deltacondist}
\mathrm{d_{rootED}}(G,G') = \sqrt{\sum_{i,j} (\sqrt{S_{i,j}} - \sqrt{S'_{i,j})}^2}.
\end{eqnarray}

When used instead of the Euclidean distance, $\mathrm{d_{rootED}}(G,G')$ may even detect small changes in the graphs. The fast belief propagation matrix $S$ is defined as
\begin{eqnarray}
\label{eq:deltaconS}
S = [I+ \varepsilon^2 D - \varepsilon A]^{-1},
\end{eqnarray}
where $\varepsilon=1/(1 + max_i D_{ii})$ and it is assumed to be $\varepsilon \ll 1$, so that S can be rewritten in a matrix power series as:
\begin{eqnarray}
\label{eq:deltaconS_approx}
S \approx I + \varepsilon A + \varepsilon^2 (A^2-D) + \dots.
\end{eqnarray}
Fast belief propagation is a fast algorithm and is designed to perceive both global and local structures of the graph.

\subsection*{Criminal networks data sources}

Our analysis focuses on nine real criminal networks of different nature (see Table~\ref{tab:table1}). The first six networks relate to three distinct Mafia operations, while the other three are linked to street gangs and terrorist organizations.

\begin{table}[!ht]
\begin{adjustwidth}{-2.25in}{0in} 
\centering
\caption{
{\bf Criminal networks characterization.}}
\begin{tabular}{|c|c|l|l|c|}
\hline
{\bf Investigation} & \multicolumn{3}{|c|}{\bf Network} & {\bf Source} \\ \cline{2-4}
& \textbf{Name} & \textbf{Nodes} & {\bf Edges} &\\ \thickhline
\hline
\makecell{Montagna Operation \\ (Sicilian Mafia) \\ 2003-2007} & \makecell{ MN \\[2pt] PC} & Suspects & \makecell[l]{Physical Surveillance \\ Audio Surveillance}& \cite{Ficara2020,Calderoni2020,Cavallaro2020, Cavallaro2021, Zenodo2020} \\ \hline
\makecell{Infinito Operation \\ (Lombardian 'Ndrangheta) \\ 2007-2009} & SN & Suspects & \makecell[l]{Physical and \\ Audio Surveillance} & \cite{Calderoni2014, Calderoni2014b, Calderoni2015, Calderoni2017, Grassi2019} \\ \hline
\makecell{Oversize Operation \\ (Calabrian 'Ndrangheta) \\ 2000-2009} & \makecell{WR \\ AW \\ JU} & Suspects & \makecell[l]{Audio Surveillance \\ Physical Surveillance \\ Audio Surveillance} & \cite{Berlusconi2016, piccardi2016} \\ \hline
\makecell{Swedish Police Operation \\ (Stockholm Street Gangs) \\ 2000-2009} & SV & Gang members & Physical Surveillance & \cite{Rostami2015, rostami_mondani_2015} \\ \hline
\makecell{Caviar Project \\ (Montreal Drug Traffickers) \\ 1994-1996} & CV & Criminals & Audio Surveillance & \cite{Morselli2008} \\ \hline
\makecell{Abu Sayyaf Group \\ (Philippines Kidnappers) \\ 1991-2011} & PK & Kidnappers & Attacks locations & \cite{Gerdes2014} \\ \hline
\end{tabular}
\label{tab:table1}
\end{adjustwidth}
\end{table}

The Montagna Operation was an investigation concluded in 2007 by the Public Prosecutor’s Office of Messina (Sicily) focused on the Sicilian Mafia groups known as Mistretta and Batanesi clans.
Between 2003 and 2007 these families infiltrated several economic activities including public works in the area, through a cartel of entrepreneurs close to the Sicilian Mafia. The main data source is the pre-trial detention order issued by the Preliminary Investigation Judge of Messina on March 14, 2007. 

The order concerned a total of 52 suspects, all charged with the crime of participation in a Mafia clan as well as other crimes such as theft, extortion or damaging followed by arson. From the analysis of this legal document we built two weighted and undirected graphs: the Meeting network (MN) with 101 nodes and 256 edges,
and the Phone Calls (PC) network with 100 nodes and 124 edges (see Table~\ref{tab:mafia}). In both networks, nodes are suspected criminals and edges represent meetings (MN), or recorded phone calls (PC). These original datasets have been already studied in some of our previous works~\cite{Ficara2020,Calderoni2020,Cavallaro2020, Cavallaro2021, Cavallaro2021IoT} and they are available on Zenodo~\cite{Zenodo2020}.

The Infinito Operation was a large law enforcement operation against 'Ndrangheta groups (i.e., groups of the Calabrian Mafia) and Milan cosche (i.e., crime families or clans) concluded by the courts of Milan and Reggio Calabria, Italian cities situated in Northern and Southern Italy, respectively. The investigation started 2003 is still in progress. On July 5, 2010, the Preliminary Investigations Judge of Milan issued a pre-trial detention order for 154 people, with charges ranging from mafia-style association to arms trafficking, extortion and intimidation for the awarding of contracts or electoral preferences.
The dataset was extracted from this judicial act and is available as a $2$-mode matrix on the UCINET~\cite{Borgatti2002} website (Link: \url{https://sites.google.com/site/ucinetsoftware/datasets/covert-networks/ndranghetamafia2}).
The Infinito Operation dataset was investigated by Calderoni and his co-authors in several works~\cite{Calderoni2014, Calderoni2014b, Calderoni2015, Calderoni2017, Grassi2019}.
From the original $2$-mode matrix, we constructed the weighted and undirected graph Summits Network (SN) with 156 nodes and 1619 edges (Table~\ref{tab:mafia}). Nodes are suspected members of the 'Ndrangheta criminal organization. Edges are summits (i.e., meetings whose purpose is to make important decisions and/or affiliations, but also to solve internal problems and to establish roles and powers) taking place between 2007 and 2009. This network describes how many summits in common any two suspects have. Attendance at summits was registered by police authorities through wiretapping and observations during this operation.

The Oversize Operation is an investigation lasting from 2000 to 2006, which targeted more than 50 suspects of the Calabrian 'Ndrangheta involved in international drug trafficking, homicides, and robberies. The trial led to the conviction of the main suspects from 5 to 22 years of imprisonment between 2007-2009. Berlusconi et al.~\cite{Berlusconi2016} studied three unweighted and undirected networks extracted from three judicial documents corresponding to three different stages of the criminal proceedings (Table~\ref{tab:mafia}): wiretap records (WR), arrest warrant (AW), and judgment (JU). 
Each of these networks has 182 nodes which corresponding to the individuals involved in illicit activities. The WR network has 247 edges which represent the wiretap conversations transcribed by the police and considered relevant at first glance. The AW network contains 189 edges which are meetings emerging from the physical surveillance. The JU network has 113 edges which are wiretap conversations emerging from the trial and several other sources of evidence, including wiretapping and audio surveillance. These datasets are available as three $1$-mode matrices on Figshare~\cite{piccardi2016}.

\begin{table}[!ht]
\begin{adjustwidth}{-2.25in}{0in} 
\centering
\caption{
{\bf Mafia networks properties.}}
\begin{tabular}{|l|l|l|l|l|l|l|l|}
\hline
\textbf{Network} & \textbf{MN} & \textbf{PC} & \textbf{SN} & \textbf{WR} & \textbf{AW} & \textbf{JU} \\ \thickhline
weights & weighted & weighted & weighted & unweighted & unweighted & unweighted\\ \hline
directionality & undirected & undirected & undirected & undirected & undirected & undirected \\ \hline
connectedness & false & false & false & false & false & false \\ \hline
n. of nodes $n$ & 101 & 100 & 156 & 182 & 182 & 182 \\ \hline
n. of isolated nodes $n_i$ & 0 & 0 & 5 & 0 & 36 & 93\\ \hline
n. of edges $m$ & 256 & 124 & 1619 & 247 & 189 & 113\\ \hline
n. of components $\lvert cc\rvert$ & 5 & 5 & 6 & 3 & 38 & 96\\ \hline
max avg. path length $\langle d\rangle$ for $cc$ & 3.309 & 3.378 & 2.361 & 3.999 & 4.426 & 3.722 \\ \hline
max shortest path length $d$ & 7 & 7 & 5 & 8 & 9 & 7 \\ \hline
density $\delta$ & 0.051 & 0.025 & 0.134 & 0.015 & 0.011 & 0.007 \\ \hline
avg. degree $\langle k \rangle$ & 5.07 & 2.48 & 20.76 & 2.71 & 2.08 & 1.24 \\ \hline
max degree $k$ & 24 & 25 & 75 & 32 & 29 & 13\\ \hline
avg. clust. coeff. $\langle C \rangle$ & 0.656 & 0.105 & 0.795 & 0.149 & 0.122 & 0.059\\ \hline
\end{tabular}
\label{tab:mafia}
\end{adjustwidth}
\end{table}

The Stockholm street gangs dataset was extracted from the National Swedish Police Intelligence (NSPI), which collects and registers the information from different kinds of intelligence sources to identify gang membership in Sweden. The organization investigated here is a Stockholm-based street gang localised in southern parts of Stockholm County, consisting of marginalised suburbs of the capital. All gang members are male with high levels of violence, thefts, robbery and drug-related crimes. Rostami and Mondani~\cite{Rostami2015} constructed the Surveillance (SV) network (Table~\ref{tab:others}). It contains data from the General Surveillance Register (GSR) which covers the period 1995–2010 and aims to facilitate access to the personal information revealed in law enforcement activities needed in police operations. SV is a weighted network with 234 nodes that are gang members. Some of them were no longer part of the gang in the period covered by the data and have been included as isolated nodes. The link weight counts the number of occurrence of a given edge. This dataset is available on Figshare~\cite{rostami_mondani_2015}.

Project Caviar~\cite{Morselli2008} was a unique investigation against hashish and cocaine importers operating out of Montreal, Canada. The network was targeted between 1994 and 1996 by a tandem investigation uniting the Montreal Police, the Royal Canadian Mounted Police, and other national and regional law-enforcement agencies from England, Spain, Italy, Brazil, Paraguay, and Colombia. In a 2-year period, 11 importation drug consignments were seized at different moments and arrests only took place at the end of the investigation. The principal data sources are the transcripts of electronically intercepted telephone conversations between suspects submitted as evidence during the trials of 22 individuals. Initially, 318 individuals were extracted because of their appearence in the surveillance data. From this pool, 208 individuals were not implicated in the trafficking operations. Most were simply named during the many transcripts of conversations, but never detected. Others who were detected had no clear participatory role within the network (e.g., family members or legitimate entrepreneurs). The final Caviar (CV) network was composed of 110 nodes. The $1$-mode matrix with weighted and directed edges is available on the UCINET~\cite{Borgatti2002} website. 
(Link: \url{https://sites.google.com/site/ucinetsoftware/datasets/covert-networks/caviar}).
From this matrix, we extracted an undirected and weighted network with 110 nodes which are criminals and 205 edges which represent the communications exchanges between them (see Table~\ref{tab:others}). Weights are level of communication activity. 

Philippines Kidnappers data refer to the Abu Sayyaf Group (ASG)~\cite{Gerdes2014}, a violent non-state actor operating in the Southern Philippines. In particular, this dataset is related to the Salast movement that has been founded by Aburajak Janjalani, a native terrorist of the Southern Philippines in 1991. ASG is active in kidnapping and other kinds of terrorist attacks. The reconstructed $2$-mode matrix is available on UCINET~\cite{Borgatti2002} (Link: \url{https://sites.google.com/site/ucinetsoftware/datasets/covert-networks/philippinekidnappings}).
From the $2$-mode matrix, we constructed a weighted and undirected graph called Philippines Kidnappers (PK) (see Table~\ref{tab:others}). The PK network has 246 nodes and 2571 edges. Nodes are terrorist kidnappers of the ASG. Edges are the terrorist events they have attended. This network describes how many events in common any two kidnappers have. 

\begin{table}[!ht]
\centering
\caption{
{\bf Street gangs and terrorist networks properties.}}
\begin{tabular}{|l|l|l|l|}
\hline
\textbf{Network} & \textbf{SV} & \textbf{CV} & \textbf{PK} \\ \thickhline
weights & weighted & weighted & weighted \\ \hline
directionality & undirected & undirected & undirected \\ \hline
connectedness & false & true & false \\ \hline
nr. of nodes $n$ & 234 & 110 & 246 \\ \hline
nr. of isolated nodes $n_i$ & 12 & 0 & 16 \\ \hline
nr. of edges $m$ & 315 & 205 & 2571 \\ \hline
nr. of components $cc$ & 13 & 1 & 26 \\ \hline
max avg. path length $\langle d\rangle$ for $cc$ & 3.534 & 2.655 & 3.034 \\ \hline
max shortest path length $d$ & 6 & 5 & 9 \\ \hline
density $\delta$ & 0.012 & 0.034 & 0.085 \\ \hline
avg. degree $\langle k \rangle$ & 2.69 & 3.73 & 20.9 \\ \hline
max degree $k$ & 34 & 60 & 78 \\ \hline
avg. clustering coeff. $\langle C \rangle$ & 0.15 & 0.335 & 0.753 \\ \hline
\end{tabular}
\label{tab:others}
\end{table}

Useful information about Mafia, street gangs and terrorist networks is provided in Tables~\ref{tab:mafia} and~\ref{tab:others}, including edges weight and directionality, connectedness, number of nodes including isolated ones, number of edges, number of connected components, maximum average path length for each connected component, maximum shortest path length, average degree, maximum degree and the average clustering coefficient. The CV network seems to be the only fully connected network (i.e., $\lvert cc \rvert = 1$) and, for this reason, in all the considered networks we chose to compute the average path length for the single components and then to show the maximum value. 

\begin{figure}[!h]
\caption{{\bf Degree Distributions.} The degree distribution $p_k$ provides the probability that a randomly selected node in each criminal network has degree $k$. Same colors imply the networks belong to the same police investigation.}
\includegraphics[width=\textwidth]{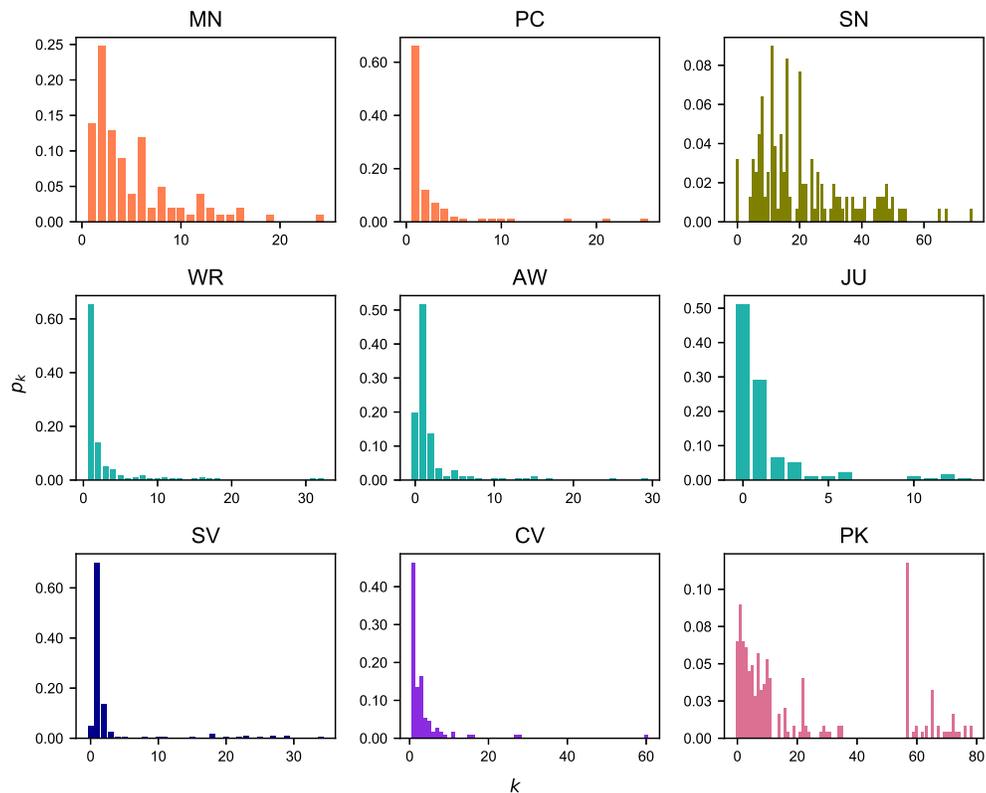}
\label{fig:degdist}
\end{figure}

Then, we showed the degree distributions for each criminal network as a normalised histogram (see Fig.~\ref{fig:degdist}). MN, PC, WR, AW, JU, SV and CV have similar degree distributions in which most nodes have a relatively small degree $k$ with values around $0$, $1$ or $2$, while a few nodes have very large degree $k$ and are connected to many other nodes. SN and PK are the only networks having different degree distributions compared to other criminal networks, as most of their nodes have large degree $k$. In particular, we note that most nodes in PK are strongly connected and have a degree $k=57$.

\subsection*{Design of Experiments}
In this section we describe the technical details in the design of the experiments conducted. To understand how much partial knowledge of a criminal network may negatively affect the investigations, we have implemented several tests. 

Since we are trying to understand how much differences can be spotted based on different types/amount of data missing, we set up the experiments by two main strategies: random edges removal and nodes removal. The first case simulates the scenario in which LEAs miss to intercept some calls or to spot sporadic meetings among suspects (i.e., due to the delays in obtaining a warrant). By nodes removal we mean that the selected nodes have been removed, jointly with their incident edges, and afterwords they have been reinserted within the networks as isolated nodes. Indeed, the second case catches the hypothesis in which some suspects cannot be intercepted. For instance, if a criminal is known to be a boss but there are not enough proofs to be investigated, then that criminal can be identified as an isolated node with no incident edges.

Note that for a better comparison among the networks, the graphs have been all considered as unweighted because both AW and JU are. Furthermore, all the suspects showed as isolated nodes of the original network have been excluded. In fact, our input parameter was the edge list of the graph, which does not take into account nodes with no incident edges.

Algorithm~\ref{alg:pseudo} shows the pseudocode of our approach. In order to obtain the subgraphs, we started from the previously described datasets; then, we converted them into graphs (i.e., $G$) and, lastly, we pruned them (i.e., $G^{'}$) according to a prefixed fraction $torem=10\%$. We opted for the 10\% because the criminal networks considered are small, as they have a total number of nodes lower than 250. Afterwards, we have computed the spectral and matrix distances $d(G,G^{'})$ between the original and the pruned graphs. Each edges removal process has been repeated a fixed number of times ($nrep=100$) and the results obtained have been averaged. Thus, the averaged distances values $\langle X \rangle$ and their standard deviations $\sigma$ have been computed. 

\begin{algorithm}
\caption{Pseudocode for computing the distances}
\label{alg:pseudo}
\begin{algorithmic}[1]
    \State Parameter configuration: $nrep$, $torem$, and $check$
    \State Read the dataset and covert it as graph $G$
    \If{$check = True$}
        \State Isolate $torem$ of nodes
        \Else 
        \State Remove $torem$ of random edges
        \EndIf
        \State Compute $S(G)$ 
        \State Compute the matrices $A(G)$, $L(G)$, $\mathcal{L}(G)$ 
        \For{\texttt{$torem$}}
            {\For{$nrep$}
                \State Create a pruned graph $G^{'}$ and  compute $S^{'}(G^{'})$
                \State Compute $\mathrm{d_{rootED}}(G,G^{'})$, $d_{A}(G,G^{'})$, $d_{L}(G,G^{'})$, and $d_{\mathcal{L}}(G,G^{'})$
            \EndFor}
        \State Compute $\langle X \rangle$, $\sigma$ $\forall$ $d(G,G^{'})$ $\in$ $nrep$  
      \EndFor
\end{algorithmic}
\end{algorithm}

\section*{Results}
\begin{figure}[!h]
\begin{center}
\caption{{\bf The removal effects of a fraction $F_e$ of edges by showing the distances $d(G, G^{'})$ between the original graphs with their pruned versions.} (A) Adjacency Spectral Distance $d_A$ (B) Laplacian Spectral Distance $d_L$ (C) Normalised Laplacian Spectral Distance $d_{\mathcal{L}}$ (D) Root Euclidean Distance $\mathrm{d_{rootED}}$.}
\includegraphics[width=\textwidth]{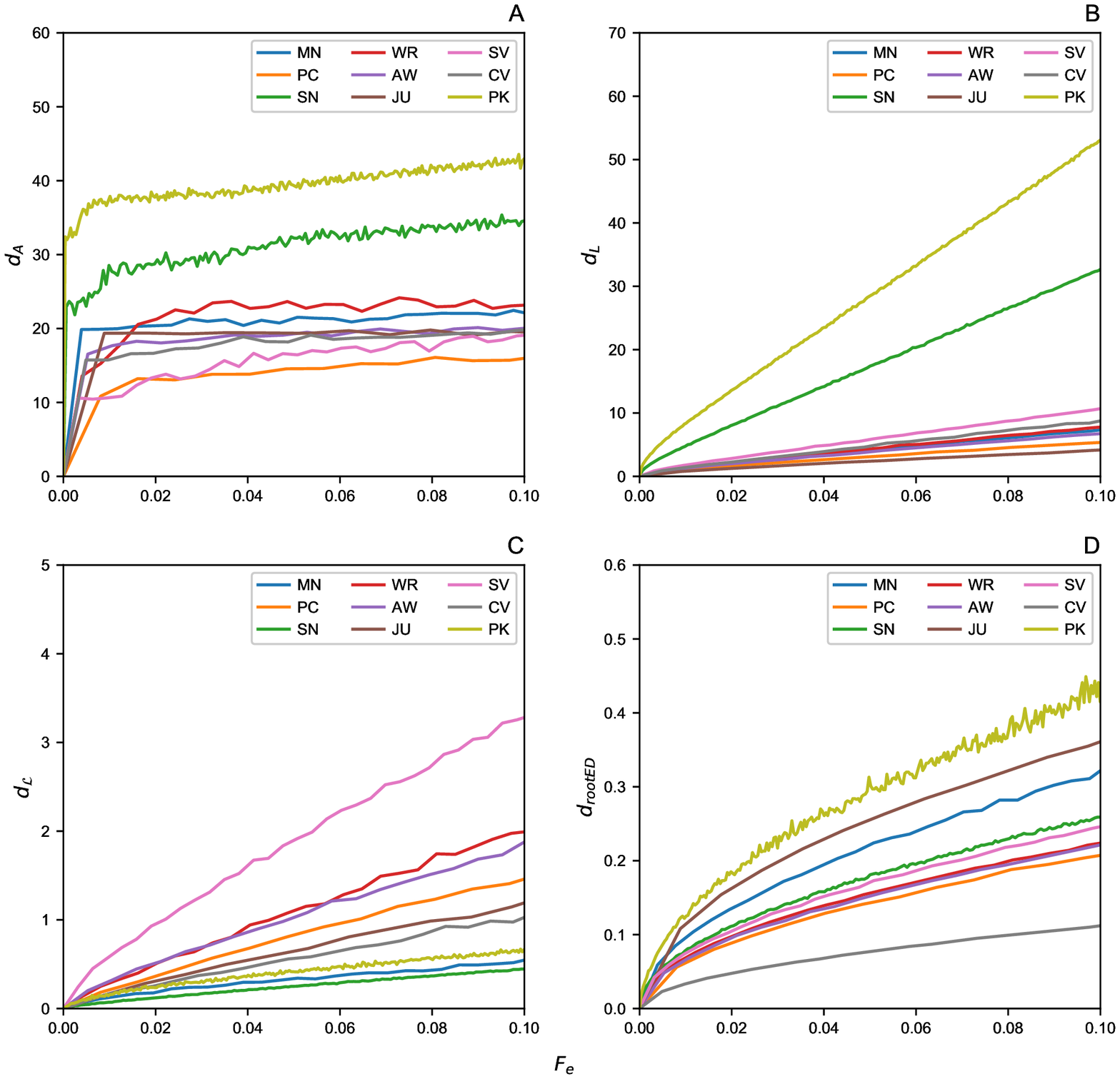}
\label{fig:edges_dist}
\end{center}
\end{figure}

\begin{figure}[!h]
\caption{{\bf The removal effects of a fraction $F_n$ of nodes by showing the distances $d(G, G^{'})$ between the original graphs with their pruned versions.} (A) Adjacency Spectral Distance $d_A$ (B) Laplacian Spectral Distance $d_L$ (C) Normalised Laplacian Spectral Distance $d_{\mathcal{L}}$ (D) Root Euclidean Distance $d_{rootED}$.} 
\includegraphics[width=\textwidth]{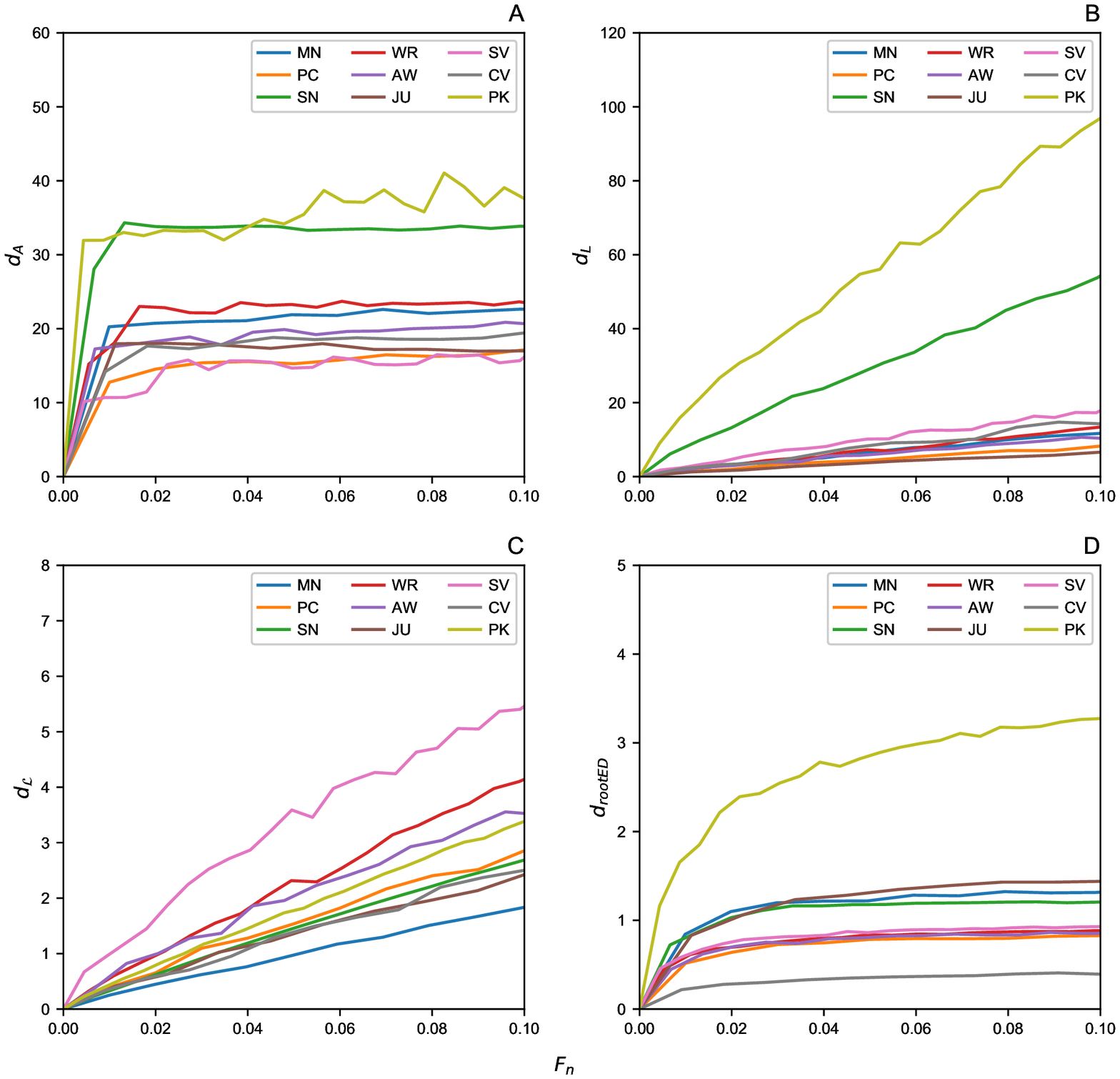}
\label{fig:nodes_dist}
\end{figure}

Here we present the results obtained from the network pruning experiments. The distance analysis between the real and the pruned networks is reported starting from the random edges removal approach (Fig.~\ref{fig:edges_dist}), moving to the analysis on the networks after node pruning (Fig.~\ref{fig:nodes_dist}). The plots show the distances between the original graphs and their pruned versions up to 10\% of edges ($F_e$) and nodes ($F_n$), respectively. 

In both removal processes, $d_A$ displays a saturation effect that makes the results difficult to be interpreted. Indeed, with a fraction of approximately the 2\% of removed elements (i.e, nodes/edges), the growth became flatter. Hence, this distance is not effective for highlighting the effects of missing data on criminal networks. Furthermore, from this metric it might seem that the two pruned networks of PK and SN show a greater deviation from their original counterparts, but this is due to the inner structure of this metric, which is highly influenced by the nodes' degree. In fact, the average degree of PK an SN (see Tables~\ref{tab:mafia} and~\ref{tab:others}) is significantly higher (i.e., $\langle k \rangle \simeq 21$) than the other networks herein studied (i.e., $1 < \langle k \rangle < 4$); moreover, their different topology is also evident from their degree distribution (see Fig.~\ref{fig:degdist}). This is the reason why these networks seem to have a more significant detachment effect than others; however, they too suffer the saturation effect mentioned above as they grow. A similar behavior has also been encountered in $d_L$ and its explanation is the same.

\begin{figure}[!h]
\caption{{\bf DeltaCon similarity $sim_{DC}$ computation} (A) Edges removal process by the fraction $F_e$ (B) Nodes removal process by the fraction $F_n$.} 
\includegraphics[width=\textwidth]{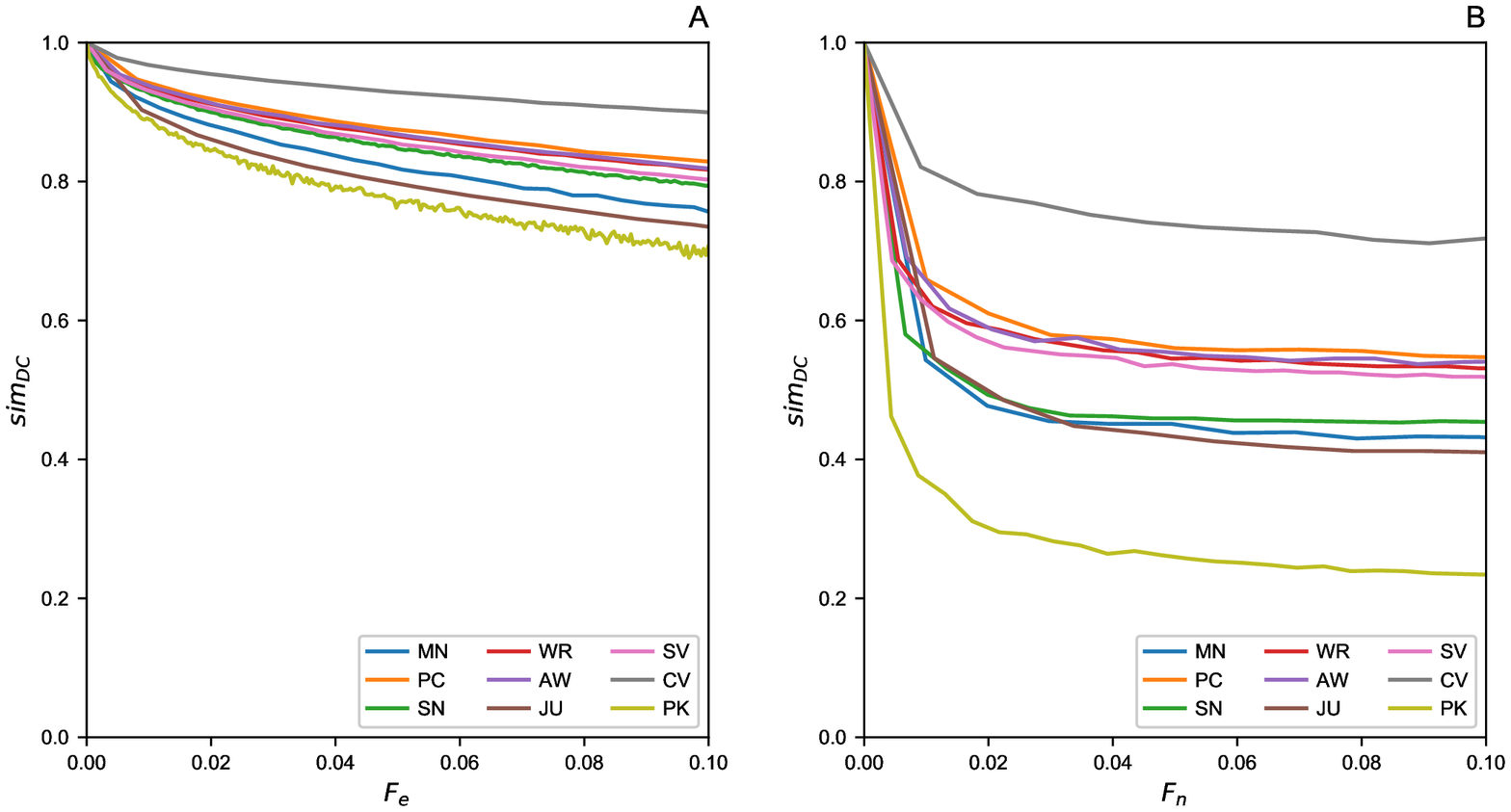}
\label{fig:sim}
\end{figure}

On the other hand, the distance metric which more effectively catches the damage caused by a significant amount of missing data is $d_{\mathcal{L}}$, where distance growth is linear. Indeed, the effects of $\langle k \rangle$ are smaller as this aspect is compressed by the structure of this distance metric. It would seem that this metric is the most effective measure compared to other spectral distances, in understanding how much lacking data affects the total knowledge of the network. A similar trend was also found in $\mathrm{d_{rootED}}$; however, for a better comparison between nodes and edges removal processes, we analysed this last metric in more detail by considering its $DeltaCon$ similarity $sim_{DC}$ (Fig.~\ref{fig:sim}).

The figure shows the difference between the original and pruned networks as the fraction of elements removed increases (i.e., $F_e$ for edges and $F_n$ for nodes). 

Before starting pruning, we have $sim_{DC}=1$. Afterwards, the drop start to became more evident as the fraction $F$ increases. In addition, as expected, the nodes removal process affects more significantly the networks. This means that if the lack of data relates to sporadically missed wiretaps, or to just a few random connections between suspects, then the network structure is not as much misinterpreted as if the case when one suspect has not been tracked at all. Indeed, pruning the network at its 2\%, causes a $sim_{DC}\geq0.8$ for edges pruning, compared with a $sim_{DC}\simeq 0.2$ for the nodes ones. Therefore, even when a small amount of suspects are not included in the investigations, this can lead to a very different network. The exclusion of the suspects could be voluntary or not. It highly depends on the overall investigation process, starting from the very preliminary analysis, and up to the judges' decision to allow warrants, or to exclude data considered irrelevant for the current investigation.

\section*{Discussion}
In this paper we analysed nine datasets of real criminal networks extracted from six police operations to investigate on the effects of missing data. More specifically, three of them rely on Mafia operations (i.e., Montagna, Infinito, and Oversize) and the remaining ones refer to other criminal networks such as street gangs, drug traffics, or terrorist networks (i.e., Stockholm street gangs, Caviar Project, Philippines Kidnappers). 

Our study focused on a careful analysis of the datasets, in order to simulate the events where some of the data is missing. In particular, two different scenarios have been considered: \begin{enumerate*}[label=(\roman*)]
\item random edges removal, which simulates the case in which LEAs miss to intercepts some calls or to spot sporadic meetings among suspects, and 
\item nodes removal that catches the hypothesis in which some suspects cannot be intercepted for some reason. For instance, if a criminal is known to be a boss, but there are not enough proofs to be investigated, then the criminal can be identified by an isolated node with no incident edges on it.
\end{enumerate*}

In order to quantify the difference between the original network and its pruned version, we computed several distance metrics, to the one which is most sensitive. Hence, we computed the Adjacency, Laplacian and Normalised Spectral distances (i.e., $d_A(G,G^{'})$, $d_L(G,G^{'})$, and $d_{\mathcal{L}}(G,G^{'})$, respectively) plus the Root Euclidean Distance (i.e., $\mathrm{d_{rootED}}(G,G^{'})$) because this metric allows to compute the DeltaCon similarity (i,e., $sim_{DC}$), which can quantify even small differences between two graphs in $[0,1]$. The pruning process involved removing up to 10\% of elements, that is the fraction $F_e$ of edges and $F_n$ of nodes. This percentage has been chosen as the networks size was quite small (less than 250 nodes per each dataset).

Our analysis suggests that \begin{enumerate*}[label=(\roman*)] \item the spectral metric $d_{\mathcal{L}}(G,G^{'})$ is best at catching the expected linear growth of differences with the incomplete graph against its complete counterpart; \item the nodes removal process is significantly more damaging than random edges removal; \end{enumerate*} thus, it translates to a negligible error in terms of graph analysis when, for example, some wiretaps are missing. Indeed, in terms of $sim_{DC}$ drop, there is a 30\% difference from the real network, for a pruned version at 10\%. On the other hand, it is crucial to be able to investigate the suspects in time because excluding them from the investigation could produce a very different network respect to the real one, that is up to 80\% of $sim_{DC}$ drop on some networks. 

A final consideration concerns the impossibility of conducting this type of analysis through the use of Machine Learning, as it is currently practically impossible to obtain a sufficient number of reliable and complete datasets of real criminal networks in order to be able to conduct an appropriate training of a Neural Network.

For the future, we plan to extend the analysis by considering weights as well. This will allow to conduct a comparative analysis of the missing data effects when not only the connections between nodes, but also their frequency is known. Another interesting aspect to be considered is the network behaviour after their pruning in both criminal and general social networks. Lastly, using the future knowledge gained from the network analysis herein presented, one could try to define an artificial network able to accurately simulate the behavior of real criminal networks.

\nolinenumbers

\bibstyle{plos2015}
\bibliography{mybib}

\begin{thebibliography}{10}

\bibitem{Finckenauer2005}
Finckenauer JO.
\newblock Problems of definition: What is organized crime?
\newblock Trends in Organized Crime. 2005;8(3):63--83.
\newblock doi:{10.1007/s12117-005-1038-4}.

\bibitem{thrasher2013gang}
Thrasher FM.
\newblock The gang: A study of 1,313 gangs in Chicago.
\newblock University of Chicago Press; 2013.

\bibitem{adler1993wheeling}
Adler PA.
\newblock Wheeling and dealing: An ethnography of an upper-level drug dealing
  and smuggling community.
\newblock Columbia University Press; 1993.

\bibitem{reuter1983disorganized}
Reuter P.
\newblock Disorganized crime: The economics of the visible hand.
\newblock MIT press Cambridge, MA; 1983.

\bibitem{Morselli2008}
Morselli C.
\newblock Inside Criminal Networks.
\newblock Studies of Organized Crime. Springer New York; 2008.

\bibitem{gambetta1996sicilian}
Gambetta D.
\newblock The Sicilian Mafia: The Business of Private Protection.
\newblock Cambridge: Harvard University Press; 1996.

\bibitem{Campana2016}
Campana P.
\newblock {Explaining criminal networks: Strategies and potential pitfalls}.
\newblock Methodological Innovations. 2016;9:205979911562274.
\newblock doi:{10.1177/2059799115622748}.

\bibitem{Johnsen2018IdentifyingCI}
Johnsen JW, Franke K.
\newblock Identifying Central Individuals in Organised Criminal Groups and
  Underground Marketplaces.
\newblock In: Shi Y, Fu H, Tian Y, Krzhizhanovskaya VV, Lees MH, Dongarra J,
  et~al., editors. Computational Science -- ICCS 2018. Cham: Springer
  International Publishing; 2018. p. 379--386.

\bibitem{Calderoni2020}
Calderoni F, Catanese S, {De Meo} P, Ficara A, Fiumara G.
\newblock {Robust link prediction in criminal networks: A case study of the
  Sicilian Mafia}.
\newblock Expert Systems with Applications. 2020;161:113666.
\newblock doi:{10.1016/j.eswa.2020.113666}.

\bibitem{morselliglance}
MORSELLI C, ROY J.
\newblock Brokerage qualifications in ringing operations.
\newblock Criminology. 2008;46:71 -- 98.
\newblock doi:{10.1111/j.1745-9125.2008.00103.x}.

\bibitem{Duijn2014}
Duijn PAC, Kashirin V, Sloot PMA.
\newblock The Relative Ineffectiveness of Criminal Network Disruption.
\newblock Scientific Reports. 2014;4(1):4238.
\newblock doi:{10.1038/srep04238}.

\bibitem{Rostami2015}
Rostami A, Mondani H.
\newblock The Complexity of Crime Network Data: A Case Study of Its
  Consequences for Crime Control and the Study of Networks.
\newblock PLOS ONE. 2015;10(3):1--20.
\newblock doi:{10.1371/journal.pone.0119309}.

\bibitem{Robinson2018}
Robinson D, Scogings C.
\newblock The detection of criminal groups in real-world fused data: using the
  graph-mining algorithm ``GraphExtract''.
\newblock Security Informatics. 2018;7(1):2.
\newblock doi:{10.1186/s13388-018-0031-9}.

\bibitem{Villani2019}
Villani S, Mosca M, Castiello M.
\newblock {A virtuous combination of structural and skill analysis to defeat
  organized crime}.
\newblock Socio-Economic Planning Sciences. 2019;65(C):51--65.
\newblock doi:{10.1016/j.seps.2018.01.00}.

\bibitem{Ficara2020}
Ficara A, Cavallaro L, De~Meo P, Fiumara G, Catanese S, Bagdasar O, et~al.
\newblock Social Network Analysis of {Sicilian Mafia} Interconnections.
\newblock In: Cherifi H, Gaito S, Mendes JF, Moro E, Rocha LM, editors. Complex
  Networks and Their Applications VIII. Cham: Springer International
  Publishing; 2020. p. 440--450.

\bibitem{Cavallaro2020}
Cavallaro L, Ficara A, {De Meo} P, Fiumara G, Catanese S, Bagdasar O, et~al.
\newblock {Disrupting resilient criminal networks through data analysis: The
  case of Sicilian Mafia}.
\newblock PLOS ONE. 2020;15(8):1--22.
\newblock doi:{10.1371/journal.pone.0236476}.

\bibitem{Cavallaro2021IoT}
Cavallaro L, Bagdasar O, De~Meo P, Fiumara G, Liotta A.
\newblock In: Fortino G, Liotta A, Gravina R, Longheu A, editors. Graph and
  Network Theory for the Analysis of Criminal Networks. Cham: Springer
  International Publishing; 2021. p. 139--156.

\bibitem{campanavarese2013}
Campana P, Federico V.
\newblock Cooperation in criminal organizations: Kinship and violence as
  credible commitments.
\newblock Rationality and Society. 2013;25:263--289.
\newblock doi:{10.1177/1043463113481202}.

\bibitem{snaresearch}
Doreian P.
\newblock Doing Social Network Research, G. Robins Sage, London (2015).
\newblock Social Networks. 2015;43.
\newblock doi:{10.1016/j.socnet.2015.04.007}.

\bibitem{SPARROW1991251}
Sparrow MK.
\newblock The application of network analysis to criminal intelligence: An
  assessment of the prospects.
\newblock Social Networks. 1991;13(3):251--274.
\newblock doi:{https://doi.org/10.1016/0378-8733(91)90008-H}.

\bibitem{Squartini2015}
Squartini T, Mastrandrea R, Garlaschelli D.
\newblock Unbiased sampling of network ensembles.
\newblock New Journal of Physics. 2015;17(2):023052.
\newblock doi:{10.1088/1367-2630/17/2/023052}.

\bibitem{Peixoto2018}
Peixoto TP.
\newblock Reconstructing Networks with Unknown and Heterogeneous Errors.
\newblock Phys Rev X. 2018;8:041011.
\newblock doi:{10.1103/PhysRevX.8.041011}.

\bibitem{Newman2018}
Newman MEJ.
\newblock Estimating network structure from unreliable measurements.
\newblock Phys Rev E. 2018;98:062321.
\newblock doi:{10.1103/PhysRevE.98.062321}.

\bibitem{Soundarajan2014}
Soundarajan S, Eliassi-Rad T, Gallagher B.
\newblock In: A Guide to Selecting a Network Similarity Method; 2014. p.
  1037--1045.

\bibitem{Emmert2016}
Emmert-Streib F, Dehmer M, Shi Y.
\newblock Fifty years of graph matching, network alignment and network
  comparison.
\newblock Information Sciences. 2016;346-347:180 -- 197.
\newblock doi:{https://doi.org/10.1016/j.ins.2016.01.074}.

\bibitem{Donnat2018}
Donnat C, Holmes S.
\newblock {Tracking network dynamics: A survey using graph distances}.
\newblock The Annals of Applied Statistics. 2018;12(2):971--1012.
\newblock doi:{10.1214/18-AOAS1176}.

\bibitem{Tantardini2019}
Tantardini M, Ieva F, Tajoli L, Piccardi C.
\newblock Comparing methods for comparing networks.
\newblock Scientific Reports. 2019;9(1):17557.
\newblock doi:{10.1038/s41598-019-53708-y}.

\bibitem{Cavallaro2021}
Cavallaro L, Ficara A, Curreri F, Fiumara G, De~Meo P, Bagdasar O, et~al.
\newblock Graph Comparison and Artificial Models for Simulating Real Criminal
  Networks.
\newblock In: Benito RM, Cherifi C, Cherifi H, Moro E, Rocha LM, Sales-Pardo M,
  editors. Complex Networks and Their Applications IX. Cham: Springer
  International Publishing; 2021. p. 286--297.

\bibitem{barabasi2016network}
Barabási AL, Pósfai M.
\newblock Network science.
\newblock Cambridge: Cambridge University Press; 2016.
\newblock Available from: \url{http://barabasi.com/networksciencebook/}.

\bibitem{Ficara2021IoT}
Ficara A, Fiumara G, De~Meo P, Liotta A.
\newblock In: Fortino G, Liotta A, Gravina R, Longheu A, editors. Correlations
  Among Game of Thieves and Other Centrality Measures in Complex Networks.
  Cham: Springer International Publishing; 2021. p. 43--62.

\bibitem{Wilson2008}
Wilson RC, Zhu P.
\newblock A study of graph spectra for comparing graphs and trees.
\newblock Pattern Recognition. 2008;41(9):2833 -- 2841.
\newblock doi:{https://doi.org/10.1016/j.patcog.2008.03.011}.

\bibitem{Wills2020}
Wills P, Meyer FG.
\newblock {Metrics for graph comparison: A practitioner's guide}.
\newblock PLOS ONE. 2020;15(2):e0228728.
\newblock doi:{10.1371/journal.pone.0228728}.

\bibitem{koutra2013}
Koutra D, Vogelstein JT, Faloutsos C.
\newblock DELTACON: A Principled Massive-Graph Similarity Function.
\newblock Proceedings of the 2013 SIAM International Conference on Data Mining.
  2013; p. 162--170.
\newblock doi:{10.1137/1.9781611972832.18}.

\bibitem{Zenodo2020}
Cavallaro L, Ficara A, De~Meo P, Fiumara G, Catanese S, Bagdasar O, et~al..
  {Criminal Network: The Sicilian Mafia. ``Montagna Operation"}; 2020.
\newblock doi:{10.5281/zenodo.3938818}.

\bibitem{Calderoni2014}
{Calderoni} F, {Piccardi} C.
\newblock Uncovering the Structure of Criminal Organizations by Community
  Analysis: The Infinito Network.
\newblock In: 2014 Tenth International Conference on Signal-Image Technology
  and Internet-Based Systems; 2014. p. 301--308.

\bibitem{Calderoni2014b}
Calderoni F.
\newblock In: Masys AJ, editor. Identifying Mafia Bosses from Meeting
  Attendance. Cham: Springer International Publishing; 2014. p. 27--48.
\newblock Available from: \url{https://doi.org/10.1007/978-3-319-04147-6_2}.

\bibitem{Calderoni2015}
Calderoni F.
\newblock In: Predicting Organized Crime Leaders; 2015. p. 89--110.
\newblock Available from: \url{http://hdl.handle.net/10807/68084}.

\bibitem{Calderoni2017}
Calderoni F, Brunetto D, Piccardi C.
\newblock Communities in criminal networks: A case study.
\newblock Social Networks. 2017;48:116--125.
\newblock doi:{https://doi.org/10.1016/j.socnet.2016.08.003}.

\bibitem{Grassi2019}
Grassi R, Calderoni F, Bianchi M, Torriero A.
\newblock Betweenness to assess leaders in criminal networks: New evidence
  using the dual projection approach.
\newblock Social Networks. 2019;56:23--32.
\newblock doi:{https://doi.org/10.1016/j.socnet.2018.08.001}.

\bibitem{Berlusconi2016}
Berlusconi G, Calderoni F, Parolini N, Verani M, Piccardi C.
\newblock Link Prediction in Criminal Networks: A Tool for Criminal
  Intelligence Analysis.
\newblock PLOS ONE. 2016;11(4):1--21.
\newblock doi:{10.1371/journal.pone.0154244}.

\bibitem{piccardi2016}
Piccardi C, Berlusconi G, Calderoni F, Parolini N, Verani M. Oversize network;
  2016.
\newblock doi:{10.6084/m9.figshare.3156067.v1}.

\bibitem{rostami_mondani_2015}
Rostami A, Mondani H. Network complexity data; 2015.
\newblock doi:{10.6084/m9.figshare.1297161.v1}.

\bibitem{Gerdes2014}
Gerdes LM, Ringler K, Autin B.
\newblock Assessing the Abu Sayyaf Group's Strategic and Learning Capacities.
\newblock Studies in Conflict \& Terrorism. 2014;37(3):267--293.
\newblock doi:{10.1080/1057610X.2014.872021}.

\bibitem{Borgatti2002}
Borgatti S, Everett M, Freeman L.
\newblock UCINET for Windows: Software for social network analysis; 2002.

\end{thebibliography}

\end{document}